# A Pinned Polymer Model of Posture Control

Carson C. Chow[1] and James J. Collins[1,2]

[1]*NeuroMuscular Research Center and* [2]*Department of Biomedical Engineering*

*Boston University, 44 Cummington Street, Boston, Massachusetts 02215*

(January 11, 1995)

## Abstract

A phenomenological model of human posture control is posited. The dynamics are modelled as an elastically pinned polymer under the influence of noise. The model accurately reproduces the two-point correlation function of experimental posture data and makes predictions for the response function of the postural control system. The physiological and clinical significance of the model is discussed.

# I. INTRODUCTION

The human postural control system is highly evolved and complex. Recent analyses of quiet-standing posture data suggest that stochastically driven dynamics may be present in the system [1,2]. Here we present a phenomenological Langevin equation that captures the underlying physical mechanisms and reproduces the experimental posture data. The model is simple enough to allow quantitative analyses. Here we obtain correlation and response functions for the model and relate the properties of these functions back to physiological origins.

In Ref. [1], Collins and De Luca measured and analyzed the time-varying displacements of the center of pressure (COP) under the feet of quietly standing subjects. They found that the dynamics of the COP time series were not consistent with low-dimensional chaos. Subsequently, they constructed the two-point correlation function in the anteroposterior (front-to-back) direction, $y$, defined by $C(t_2 - t_1) = <(y(t_2) - y(t_1))^2>$, from the data. An example is shown in Fig. 1. Collins and De Luca found that for healthy young subjects the COP two-point correlation function exhibited three approximate scaling regions where $C(\tau) \sim \tau^{2H}$ [3]. In the first region $H \approx 4/5$, in the second $H \approx 1/4$, and in the third $H \approx 0$. These regions can be described as: (1) drift-like or free streaming-like, (2) diffusive, and (3) bounded or saturated [4]. Together the above posture results suggest a continuum phenomenological stochastic model for posture control.

# II. THE MODEL

We consider the transverse motion of the human body in one spatial dimension, $y$. We would like to model the dynamics of this motion as a function of the height variable, $z$, and time, $t$. The simplest continuum model of the body is that of a flexible string or polymer under tension with friction

$$\rho \partial_t^2 y + \mu \partial_t y = T \partial_z^2 y, \tag{2.1}$$

where $\rho$ is the mass density, $\mu$ is a friction coefficient and $T$ is the tension. In upright stance, the body has an inverted pendulum-like instability which is countered by a control system. The precise form of this control system is highly complicated and depends on many physiological processes [5]. A simple approximation is to consider the body to be "pinned" to the upright position by an elastic force. In terms of the model, this would amount to embedding the polymer in an elastic sheet. The corrections of the control system cannot always be perfect [6] and this is modelled by a stochastic force. The entire postural control system is reduced to an elastically pinned polymer under the influence of stochastic fluctuations. The full equation takes the form

$$\rho\partial_t^2 y + \mu\partial_t y = T\partial_z^2 y - Ky + F(z,t), \tag{2.2}$$

where $K$ is the elastic restoring constant of the pinning and $F$ is the stochastic driving force.

Equation (2.2) can be put into a more convenient form for analysis by dividing through by $\mu$ and relabelling the constants:

$$\beta\partial_t^2 y + \partial_t y = \nu\partial_z^2 y - \alpha y + \eta(z,t). \tag{2.3}$$

In this form, $\beta$ and $\alpha^{-1}$ have dimensions of time, and $\nu$ has the dimensions of length squared divided by time. By rescaling the length, the parameter $\nu$ can also be eliminated; however, in order to keep length and time dimensions explicit, it is retained. The free parameters in the system are the two time scales $\beta$ and $\alpha^{-1}$ and those associated with the stochastic forcing.

The stochastic forcing $\eta(z,t)$ is assumed to have a correlator of

$$<\eta(z',t')\eta(z,t)> = D(t'-t)\delta(z'-z). \tag{2.4}$$

It is uncorrelated along $z$ but can possess temporal correlations represented by the function $D(\tau)$. In Fourier space we assume the correlator can be approximated by the form

$$<\eta(k,\omega)\eta(k',\omega')> = 2D\frac{\delta^{2\gamma}}{(\omega^2+\delta^2)^{\gamma}}(2\pi)^2\delta(\omega+\omega')\delta(k+k'). \tag{2.5}$$

This assumes that for long times ($t \gg \delta^{-1}$) the forcing is uncorrelated but for short times it has some power-law scaling, $<\eta(\omega')\eta(\omega)> \propto \omega^{-2\gamma}\delta(\omega'+\omega)$. These short-time correlations are assumed to arise from neuromuscular effects, such as intrinsic muscle force fluctuations and/or feedback-loop delays. The values of the parameters will be constrained by the data.

The dynamics of the model are to be compared to the measured motion of the COP. It is assumed that this motion can be effectively mimicked by the motion of a particular point along the polymer (body). Hence, it is necessary to compute the two-point correlation function $C(t_2 - t_1) = <(y(z_0,t_2) - y(z_0,t_1))^2>$ taken at a specific point $z = z_0$. In actual posture control for quiet standing, the feet are essentially fixed to the floor and hence the pinning strength is infinite there. As one goes upwards, the pinning strength is reduced. The analysis of a model with inhomogeneous pinning, however, is difficult. Nonetheless, since only the motion at one point along the body is required, we can consider a model with an "averaged" constant pinning strength. We also assume that the boundary effects are unimportant and consider an infinitely long polymer. In this case, every point along the polymer will have equivalent dynamics. These approximations are valid for strong pinning and for the dynamics near the center of mass away from the boundaries which we assume is a location that effectively mimics the COP dynamics.

The correlation function can be computed from Eq. (2.3). It is most convenient to analyze the problem in the Fourier domain. Fourier transforming Eq. (2.3) in space and time and solving for $y(k,\omega)$ yields

$$y(k,\omega) = \frac{\eta(k,\omega)}{-\beta\omega^2 + \nu k^2 + \alpha - i\omega}. \tag{2.6}$$

The auto-correlation function is then formed to yield

$$<y(k,\omega)y(k',\omega')> = \frac{2D\delta^{2\gamma}(2\pi)^2\delta(\omega+\omega')\delta(k+k')}{((\beta\omega^2 - \nu k^2 - \alpha)^2 + \omega^2)(\omega^2+\delta^2)^\gamma} \tag{2.7}$$

where Eq. (2.5) has been applied.

Inverse Fourier transforming in $\omega'$, $k'$ and $k$ yields

$$<y(z,\omega)y(z',\omega)> = 2D\delta^{2\gamma}\int_F \frac{dk}{2\pi}\frac{e^{ik(z-z')}}{[(\beta\omega^2 - \nu k^2 - \alpha)^2 + \omega^2](\omega^2+\delta^2)^\gamma}, \tag{2.8}$$

where the integration is along the contour $F$. We want the single-point temporal correlations so we set $z = z' = z_0$ in Eq. (2.8) and obtain

$$S(\omega) \equiv <y(z_0,\omega)y(z_0,\omega)> = \frac{D\delta^{2\gamma}}{\pi} \int_F \frac{1}{((\beta\omega^2 - \nu k^2 - \alpha)^2 + \omega^2)(\omega^2 + \delta^2)^\gamma} dk. \quad (2.9)$$

The two-point correlation function is given by

$$C(\tau) = 2(S(0) - S(\tau)) \quad (2.10)$$

where $S(\tau)$ is the temporal auto-correlation function obtained by inverse Fourier transforming Eq. (2.9).

The different scaling regions in frequency and hence time can be obtained by taking suitable limits of Eq. (2.9). Note that $\nu k^2/\omega$, $\beta\omega$ and $\alpha/\omega$ are dimensionless quantities. With this in mind Eq. (2.9) can be rewritten as

$$S(\omega) = \frac{D}{\pi} \frac{1}{\omega^2(\omega^2 + \delta^2)^\gamma} \int_F \frac{\delta^{2\gamma}}{((\beta\omega - \alpha/\omega - \nu k^2/\omega)^2 + 1)} dk. \quad (2.11)$$

Consider the range of parameters where $\beta^{-1} \sim \delta >> \omega >> \alpha$. Then Eq. (2.11) can be approximated by

$$S(\omega) \sim \frac{D}{\pi \nu^{1/2} \omega^{3/2}} \int_{-\infty}^{\infty} \frac{1}{u^4 + 1} du, \quad (2.12)$$

where $u = \sqrt{\nu/\omega}k$. The integral is convergent so in this regime $S(\omega) \propto \omega^{-3/2}$. By dimensional analysis, this translates to a scaling behavior of $S(\tau) \sim \tau^{1/2}$, in the time domain [7]. This corresponds to the middle scaling region (where $H \approx 1/4$) in the COP data of Fig. 1. This scaling region is valid in the range $\beta << \tau << \alpha^{-1}$.

Now consider the long-time region where $\omega << \alpha << \beta^{-1} \sim \delta$. Equation (2.9) then takes the form

$$S(\omega) \sim \frac{D}{\pi} \int_{-\infty}^{\infty} \frac{1}{(\nu k^2 + \alpha)^2} dk. \quad (2.13)$$

The integral in Eq. (2.13) is convergent so we find that $S(\omega) \propto \omega^0$, which gives $S(\tau) \sim 0$ for $\tau >> \alpha^{-1}$. This corresponds to the saturated region (where $H \approx 0$) in the COP data in Fig. 1.

Finally, in the short-time region $\omega >> \beta^{-1} >> 1$, $\delta \sim \beta$ we have $\omega^2 >> \alpha$ and $\omega^2 \beta >> 1$ which leads to

$$S(\omega) \sim \frac{D\delta^{2\gamma}}{\pi \nu^{1/2} \beta^{3/2} \omega^{3+2\gamma}} \int_{-\infty}^{\infty} \frac{1}{(1-v^2)^2 + \epsilon^2} du \tag{2.14}$$

where

$$v = \sqrt{\frac{\nu}{\beta}} \frac{k}{\omega}, \qquad \epsilon = \frac{1}{\beta \omega}. \tag{2.15}$$

In the limit $\omega \to \infty$ ($\epsilon \to 0$), the integral in Eq. (2.14) is proportional to $\epsilon^{-1}$ so $S(\omega)$ behaves as $\omega^{-2(1+\gamma)}$. By dimensional arguments we find in the time domain, $S(\tau) \sim \tau^{(1+\gamma)}$.

The behavior of the complete $S(\omega)$ is shown in Fig. 2. The calculated two-point correlation function takes the form corresponding to the COP data in Fig. 1. Fitting the parameters to the data in Fig. 1 yields values of $\alpha^{-1} \approx 10$ s, $\beta \approx 0.5$ s, and $\gamma \approx 0.72$.

The saturation width $C(\tau = \infty)$ can also be estimated. This value represents the maximum transverse extent the body attains during quiet standing. From Eq. (2.10) we see that $C(\infty) = 2S(\tau = 0)$ since $S(\tau = \infty) = 0$. Hence $C(\infty) = 2S(\tau = 0) = 2 \int_{-\infty}^{\infty} S(\omega) d\omega$. For simplicity consider $\gamma = 0$, i.e., no correlations in the noise. The temporal correlations in the system noise in healthy subjects only extend over short time periods and do not play a role in the saturation width. By making the substitution $u = \sqrt{\beta/\alpha}\omega$ and $v = \sqrt{\nu/\alpha}k$ in Eq. (2.9) we obtain

$$S(\tau = 0) = \frac{D}{\alpha\sqrt{\nu\beta}} \int_{-\infty}^{\infty} \int_{-\infty}^{\infty} \frac{dudv}{(u^2 - v^2 - 1)^2 + u^2/\alpha\beta} \tag{2.16}$$

$$\sim \frac{D}{\sqrt{\nu\alpha}} \int_{-\infty}^{\infty} \frac{dv}{(v^2+1)^{3/2}}, \tag{2.17}$$

i.e., $C(\infty) \propto D/\sqrt{\nu\alpha}$. By absorbing the constants into $D$, an effective noise amplitude $D_{\text{eff}} = C(\infty)\sqrt{\alpha}$ can be defined. (Recall that the tension parameter $\nu$ is not independent and could have been scaled out into the noise amplitude from the outset.) From the COP data in Fig. 1, we get $D_{\text{eff}} \approx 60$ mm$^2$s$^{-1/2}$.

The values of the respective parameters have implications for the postural control system. The non-zero value of $\gamma$ implies short-time correlations in the noise. This corroborates the open-loop control hypothesis in Refs [1,2]. In these short times, the "kicks" produced by the system are correlated to each other and are not reacting to the position of the COP. The body responds inertially to these kicks, resulting in drift-like dynamics. The second scaling region of $H \approx 1/4$ represents "diffusive" scaling. In this region, the dynamics involve a balance between the damping, tension, and uncorrelated noise of the system. Physically this implies that the body is aligning itself independently of whether or not it is upright. In the saturated region, the pinning is finally felt and the body then attempts to return itself to an upright position. The noise amplitude does not affect the scaling regions, but it does establish the saturation width. The physiological significance and relevance of these and different parameters are discussed in Sec. IV.

## III. RESPONSE TO PERTURBATIONS

With the pinned polymer model, the response to a perturbing force can also be calculated. From a physiological standpoint, this is important given that a significant number of contemporary investigations in posture control have involved analyses of the measured response of the human body to various external perturbations [8]. This issue is discussed further in Sec. IV.

To obtain the response function consider the polymer model Eq. (2.3) with an imposed perturbation

$$\beta \partial_t^2 y + \partial_t y = \nu \partial_z^2 y - \alpha y + \eta(z,t) + \tilde{\epsilon}(z,t), \tag{3.1}$$

where $\tilde{\epsilon}(z,t)$ is a spatiotemporal perturbation. For simplicity we consider

$$\tilde{\epsilon}(z,t) = \epsilon \delta(z)\delta(t). \tag{3.2}$$

This corresponds to an impulsive kick at $z = 0$. We Fourier-Laplace transform Eq. (3.1) and obtain

$$-\omega^2 \beta y - i\omega y = -\nu k^2 y - \alpha y + \hat{\eta}(\omega, k) + \epsilon. \tag{3.3}$$

The response function is defined as

$$g(k,\omega) = \left\langle \frac{\delta y}{\delta \epsilon} \right\rangle = \frac{1}{-\omega^2 \beta - i\omega + \nu k^2 + \alpha}. \tag{3.4}$$

Inverse transforming leads to

$$g(z,t) = \int_L \frac{d\omega}{2\pi} e^{-i\omega t} \int_{-\infty}^{\infty} \frac{dk}{2\pi} e^{ikt} \frac{1}{-\omega^2 \beta - i\omega + \nu k^2 + \alpha}. \tag{3.5}$$

The $k$ integral can be performed to yield

$$g(z,t) = \frac{1}{4\pi\sqrt{\nu\beta}} \int_L d\omega e^{-i\omega t} \frac{e^{-|z|\sqrt{(-\beta\omega^2 - i\omega + \alpha)/\nu}}}{\sqrt{-\beta\omega^2 - i\omega + \alpha}}. \tag{3.6}$$

By making the transformation $\omega \to \omega - i/2\beta$, the response function takes the form

$$g(z,t) = \frac{e^{-t/2\beta}}{4\pi\sqrt{\nu\beta}} \int_L d\omega e^{-i\omega t} \frac{e^{-Z\sqrt{a^2 - \omega^2}}}{\sqrt{a^2 - \omega^2}}, \tag{3.7}$$

where

$$a = \frac{\sqrt{4\alpha\beta - 1}}{2\beta}, \qquad Z = \sqrt{\frac{\beta}{\nu}}|z|. \tag{3.8}$$

For $4\alpha\beta > 1$, there are branch points at $\omega = \pm a$. By definition there is no response before $t < Z$ so the Laplace contour must be above the branch cut from $-a$ to $a$. For $t > Z$, we can close the contour in the lower half plane and collapse the contour around the branch cut. The path of integration is then taken along the two sides of the cut. The integral is completed with the substitution $\omega = a\sin\phi$ and some algebraic manipulations to yield

$$g(z,t) = \theta(t - Z) \frac{e^{-t/2\beta}}{2\sqrt{\nu\beta}} \frac{1}{2\pi} \int_0^{2\pi} e^{ia\sqrt{t^2 - Z^2}\sin\phi} d\phi, \tag{3.9}$$

where $\theta(\tau)$ is the step function. The integral in Eq. (3.9) is the integral representation for the zeroth-order Bessel function so

$$g(z,t) = \theta(t - Z) \frac{e^{-t/2\beta}}{2\sqrt{\nu\beta}} J_0(a\sqrt{t^2 - Z^2}). \tag{3.10}$$

For the case where $4\alpha\beta < 1$, the branch points lie along the imaginary axis. A rotation in the complex plane by $\pi/2$ in Eq. (3.7) leads to the result

$$g(z,t) = \theta(t-Z)\frac{e^{-t/2\beta}}{2\sqrt{\nu\beta}}I_0(|a|\sqrt{t^2-Z^2}), \qquad (3.11)$$

where $I_0$ is the zeroth-order modified Bessel function. At the critical value of $4\alpha\beta = 1$, the response is given by

$$g(z,t) = \theta(t-Z)\frac{e^{-t/2\beta}}{2\sqrt{\nu\beta}}. \qquad (3.12)$$

If the response is measured at the location of the perturbing kick ($z = 0$), then we can set $z = 0$ in the above expressions. The response for different values of $\beta$ and $\alpha$ for $z = 0$ are shown in Fig. 3. For small values of $a$, the response time is on the order of $2\beta$. If $\beta$ is very small, $|a|$ becomes very large and the asymptotic form of the modified Bessel function can be used, i.e., $I_0(x) \sim e^x/\sqrt{2\pi x}$ for $x \to \infty$. This then leads to the expected damped-diffusive response behavior of $g(0,t) \sim e^{-\alpha t}/\sqrt{4\pi\nu t}$. For near-zero pinning ($\alpha \sim 0$), we obtain a power-law response behavior of $g(0,t) \sim t^{-1/2}$.

## IV. DISCUSSION

With the pinned polymer model, posture control is characterized in terms of four parameters: (1) the onset-of-damping time scale $\beta$, (2) the onset-of-pinning time scale $\alpha^{-1}$, (3) the short-time noise correlation exponent $\gamma$, and (4) the effective noise amplitude $D_{\text{eff}}$. These parameters could be used to form the basis of a physiological and clinical classification scheme. For instance, deviations in a parameter or set of parameters from those for a population of age-matched, healthy individuals could serve to indicate a possible balance disorder. Each parameter has some physiological implication and so the model can make predictions of how various physiological changes (e.g., due to injury or disease) would alter the COP data.

For the quiet-standing posture data of healthy subjects [1,2], nonlinearities are not necessary — it appears that the pinning term is strong enough to negate any nonlinear effects. However, this may not be the case for certain patients with balance disorders. For instance, nonlinearities will certainly play a role for large transverse excursions. Obvious effects include stepping or falling. These catastrophic effects lie outside the realm of the model. Weak nonlinear effects could be incorporated by allowing the pinning coefficient $\alpha$ to be a function of $y$. This would be manifested as terms of the form $y^n$ in Eq. (2.3). The $y \to -y$ symmetry of Eq. (2.3) could be broken for even $n$. This would make sense from a biomechanical standpoint since anteroposterior movements are intrinsically asymmetric [9].

The quiet-standing posture data [1,2] showed that short-term correlated noise is present in the postural control system, i.e., for short times the kicks produced by the system are temporally correlated. This finding was confirmed by the model, i.e., the stochastic forcing needed to have some power-law scaling over short times in order for the model's output to reproduce the experimental posture data. These results imply that during undisturbed stance the postural feedback mechanisms are not necessarily operational over short periods of time ($\tau \lesssim 1$ s). This is in line with the open-loop control hypothesis of Collins and De Luca [1,2]. Physiologically, this scheme could arise as a result of inherent noise, feedback-loop delays, feedback thresholds, and/or sensory "dead-zones".

A discrete version of this model would involve a chain of coupled and bounded random walkers with short-time correlations. The coupling between the walkers is required for the diffusive region. The pinned polymer model considers an infinite chain but a finite number of coupled random walkers would probably suffice. The correlations could arise from time-delays in the stepping probabilities of the random walkers [10]. The advantage of a continuum version is the ease in which calculations can be made.

Various behaviors could arise for different parameter regimes. For instance, if the damping and pinning time scales were roughly equal ($\beta \approx \alpha^{-1}$), then the diffusive region would disappear. The COP would essentially drift until it was corrected by the elastic pinning. It is possible that a control strategy involving such dynamics would be advantageous for individuals who require quick movements (e.g., athletes) since the dynamics imply that the damping of the overall system is relatively weak. An individual utilizing such a strategy would be metastable and capable of making quick adjustments.

From the standpoint of the model, balance problems could be manifested in several ways. For instance, if the pinning strength were small, the COP would fluctuate diffusively until its transverse displacement exceeded the stability (balance-maintenance) threshold — this could result in a fall. If the damping were also weak so that $\beta > \alpha^{-1}$, then oscillations would occur before saturation. Falls could be triggered if the amplitude of these oscillations was large enough to exceed the stability threshold. Similarly, if the noise amplitude was sufficiently large, then the amplitude of saturation could exceed the stability threshold — as with the aforementioned scenarios, this effect could trigger a fall. From the above discussion, it is clear that the pinned polymer model may be useful for extracting clinically relevant information from quiet-standing COP time series.

The pinned polymer model has significant utility for understanding posture control quantitatively. As noted above, the model can completely characterize the statistical behavior of quiet-standing COP trajectories in terms of four parameters. Once these parameters are fixed from the "quasi-static" posture data, the model can be used to make a prediction of the dynamic response of the postural control system. The model thus enables a direct comparison between the "quasi-static" and "dynamic" postural control systems. From a physiological standpoint, this is important because it allows one to test the hypothesis that the two control modes are equivalent, i.e., under quiet-standing and dynamic conditions, the postural control system utilizes the same mechanisms. This hypothesis has not been addressed previously because of the lack of a suitable quantitative model or approach.

Experiments are under way to measure the dynamic response of the human postural control system. If the parameters from the quiet-standing posture data match those of the response function, this would support the aforementioned hypothesis and further validate the pinned polymer model. It is expected that the dynamic response will delineate when the two mechanisms differ. For large perturbing kicks, it is possible that nonlinear effects will appear and cause a deviation from linear response.

The model does not address the actual neuromuscular control mechanisms involved in balance regulation. For instance, it is unknown whether the stochastic forcing is intrinsic in the output of skeletal muscles or whether it arises from a combination of incomplete sensory information, feedback-loop delays, and/or feedback thresholds. In order to address these issues, a detailed model of posture control that accounts for neurophysiological components and body mechanics is required. However, the pinned polymer model does allow a quantitative characterization of posture control and an assessment of physiological consequences. Moreover, any detailed model must reduce to the pinned polymer model at a coarse-grained level.

## V. ACKNOWLEDGMENTS

We thank T. Hwa, G. Taga and C.J. De Luca for informative discussions. We also thank A. Pavlik for assistance with the preparation of figures. This work was supported by the National Science Foundation and the Rehab R&D Service of the Department of Veterans Affairs.

# REFERENCES

[1] J.J. Collins and C.J. De Luca, Phys. Rev. Lett. **73**, 764 (1994).

[2] J.J. Collins and C.J. De Luca, Exp. Brain Res. **95**, 308 (1993); CHAOS (in press).

[3] The results for the COP two-point correlation function in the mediolateral (side-to-side) direction were similar. (See Ref. [2].)

[4] For the subjects in Ref. [1], the scaling exponent for the first region had a mean of $0.83 \pm 0.04$, and the scaling exponent for the second region had a mean of $0.26 \pm 0.06$. Statistics were not available for the third region because the COP time series considered were not long enough to characterize this region reliably in all subjects. However, in the subjects who saturated at relatively short time intervals the scaling exponent for the third region was observed to be approximately zero.

[5] V. Dietz, Physiol. Rev. **72**, 33 (1992); J. Massion, Prog. Neurobiol. **38**, 35 (1992).

[6] C.J. De Luca *et al.*, J. Physiol. **329**, 129 (1982); K.M. Newell and D.M. Corcos (eds), *Variability and Motor Control* (Human Kinetics Publishers, Champaign, IL, 1993).

[7] Note that the actual inverse Fourier transform would involve constants that are possibly divergent but these are removed when constructing the two-point correlation function from the auto-correlation function.

[8] H.C. Diener, F.B. Horak, and L.M. Nashner, J. Neurophysiol. **59**, 1888 (1988); S.P. Moore *et al.*, Exp. Brain Res. **73**, 648 (1988); M.H. Woollacott, C. von Hosten, and B. Rösblad, Exp. Brain Res. **72**, 593 (1988); F.B. Horak, L.M. Nashner, and H.C. Diener, Exp. Brain Res. **82**, 167 (1990); H.C. Diener *et al.*, Exp. Brain Res. **84**, 219 (1991); V. Dietz *et al.*, Neurosci. Lett. **126**, 71 (1991).

[9] The present model only considers one-dimensional transverse motion. Presumably there could be coupling between the two planes of motion through torsional forces acting at various joints (e.g., ankle, knee, hip). These effects will be investigated in the future.

[10] T. Ohira and J.G. Milton (unpublished).

FIGURES

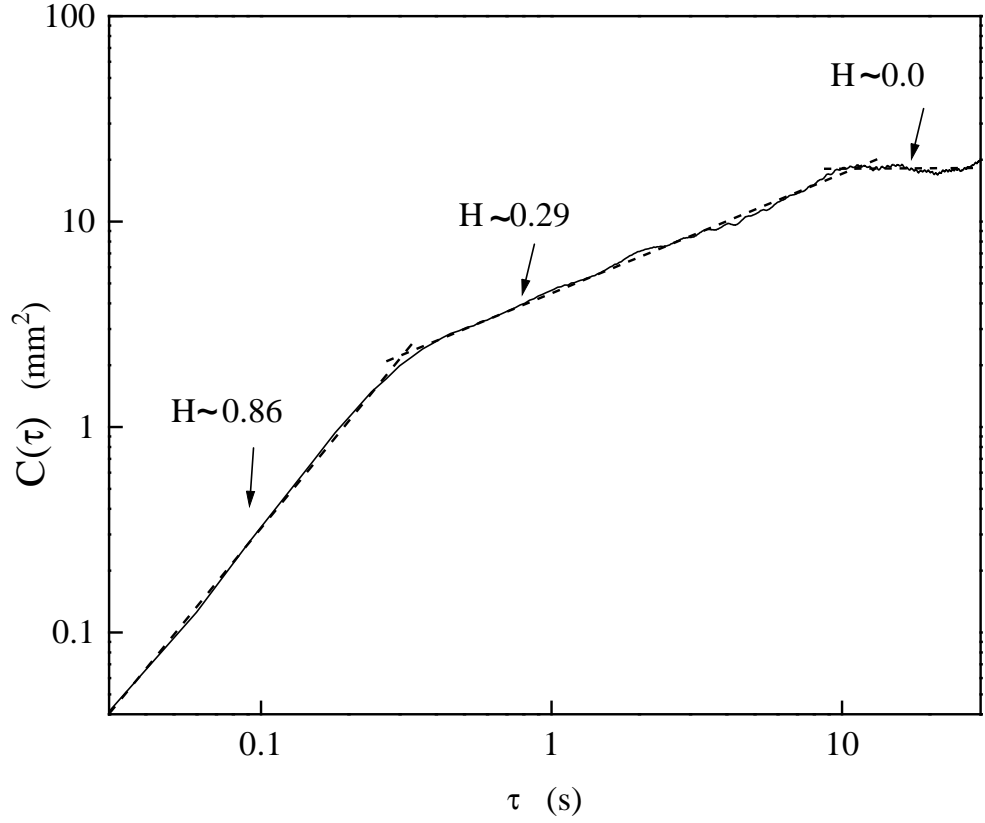

FIG. 1. A resultant log-log plot (solid line) of the COP two-point correlation function ($C(\tau)$, where $\tau$ is time interval) for a healthy young subject. This plot was generated from five 90-s COP time series. Shown also are the fitted regression lines (dashed lines) and the computed values of the scaling exponents ($H$) for the respective scaling regions.

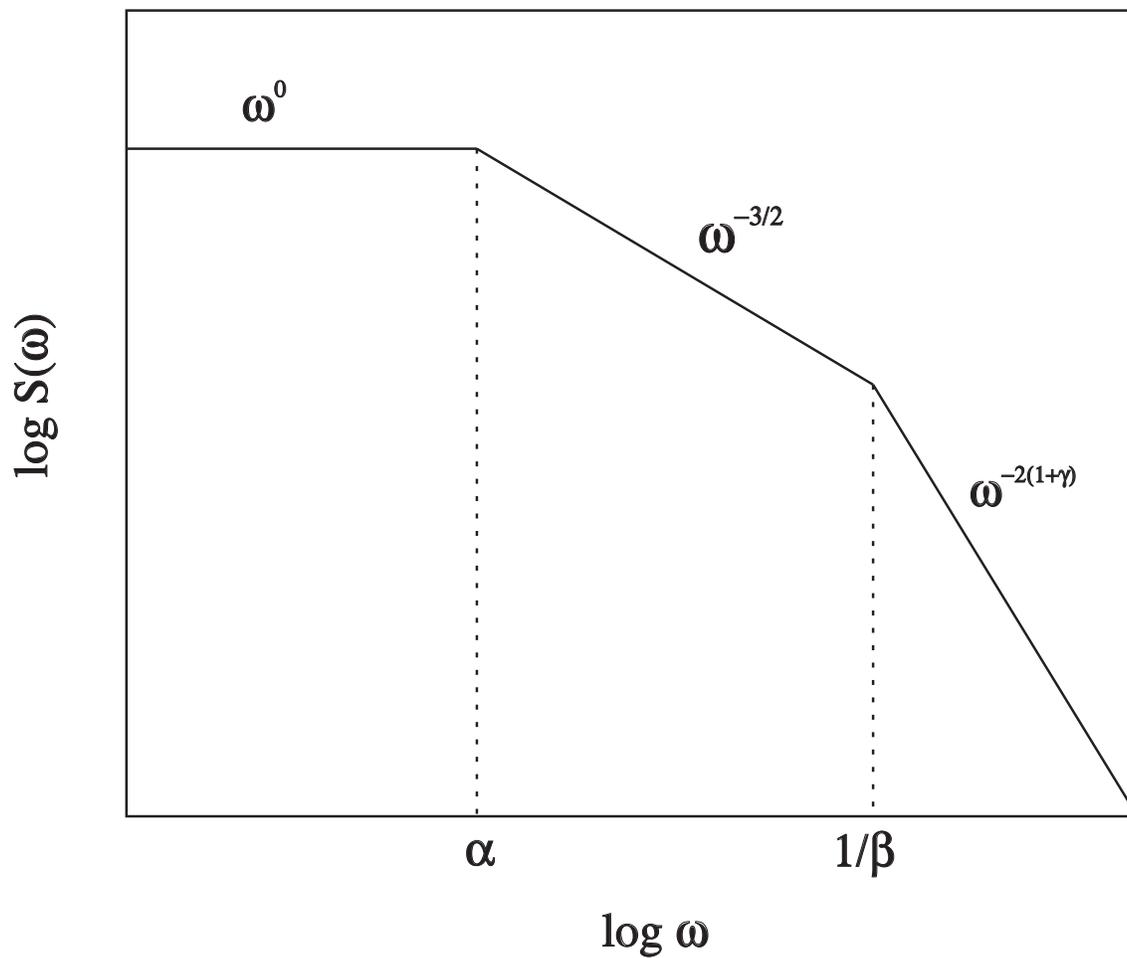

FIG. 2. The form of the auto-correlation function $S(\omega, z=0)$ in the frequency domain for the pinned polymer model.

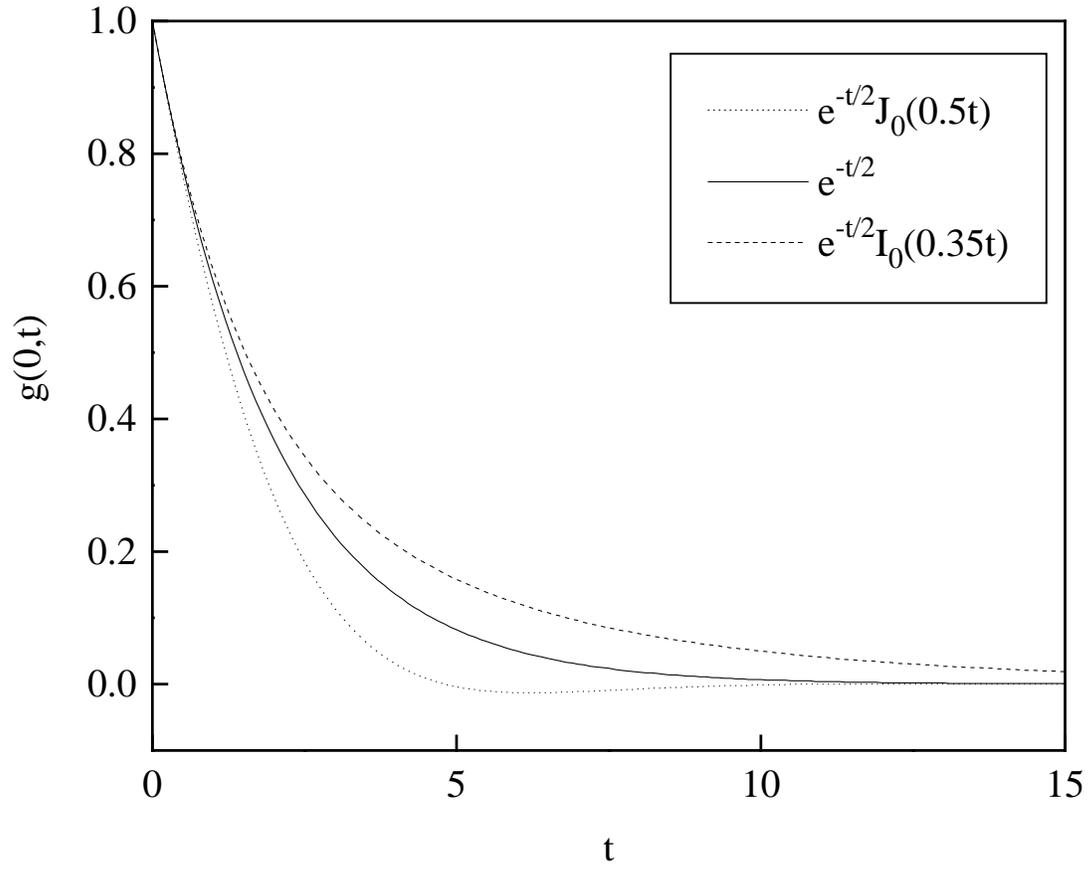

FIG. 3. Three example response functions corresponding to the parameters: $\nu = 0.25$, $\beta = 1$ and $\alpha = 0.5$ (dotted line), $\alpha = 0.25$ (solid line), $\alpha = 0.128$ (dashed line).